\documentclass[aps,twocolumn,showpacs]{revtex4}
\usepackage{dcolumn}
\usepackage{float}
\usepackage{graphicx}
\usepackage{amsmath}
\usepackage{verbatim}
\usepackage{amssymb}
\usepackage{psfrag}
\usepackage{color}
\usepackage{bm}
\usepackage{wrapfig}
\usepackage{subfigure}
\usepackage{makeidx}
\usepackage{bm}
\usepackage{epsf}
\usepackage{multirow}
\usepackage{mathrsfs}
\usepackage{braket}
\usepackage[colorlinks,linkcolor=blue,urlcolor=blue,citecolor=blue]{hyperref}

\begin{document}
\title{Varied Branches of Nondegenerate Vector Solitons}
\author{Yu-Hao Wang$^{1}$}
\author{Liang Duan$^{1}$}
\author{Yan-Hong Qin$^{2,3}$}
\author{Li-Chen Zhao$^{1,5,6}$}\email{zhaolichen3@nwu.edu.cn}

\address{$^{1}$School of Physics, Northwest University, Xi'an, 710127, China}
\address{$^{2}$College of Mathematics and System Sciences, Xinjiang University, Urumqi, China, 830046}
\address{$^{3}$Institute of Mathematics and Physics, Xinjiang University, Urumqi, China, 830046}
\address{$^{5}$NSFC-SPTP Peng Huanwu Center for Fundamental Theory, Xi'an 710127, China}
\address{$^{6}$Shaanxi Key Laboratory for Theoretical Physics Frontiers, Xi'an 710127, China}

\begin{abstract}
Our study on nondegenerate dark-bright-bright solitons in a three-component Manakov model with repulsive interactions reveals the existence of diverse branches of nondegenerate vector solitons. For fixed bright component particle numbers and a given soliton velocity, the nondegenerate dark-bright-bright solitons exhibit four distinct branches with different density profiles and phase distributions, comprising two positive mass branches and two negative mass branches. The energy-velocity dispersion relation of each pair of positive- and one negative-mass branches form a closed loop, resulting in two mutually independent loops for the soliton's overall dispersion. All soliton branches share a common maximal velocity, which is determined by the larger bright soliton particle number. Linear stability analysis shows that all these branches are stable against weak perturbations. Extending to an $N$-component Manakov system, the nondegenerate solitons have $2^{N-1}$ distinct branches, of which $2^{N-2}$ branches solitons is positive mass and $2^{N-2}$ branches solitons is negative mass. Each pair of positive- and negative-mass branches form a closed dispersion relation loop, so that the vector solitons have $2^{N-2}$ disjoint loops. These results uncover the rich branches and interesting dispersion relations of nondegenerate vector solitons in multi-component models.
\end{abstract}

\date{\today}
\pacs{05.45.Yv, 03.75.Lm, 03.75.Mn, 02.30.Ik}
\maketitle

\section{Introduction}
Matter wave solitons are collective excitations in Bose-Einstein condensates (BECs) that exhibit particle-like behavior. Their center-of-mass motion is dictated by the dispersion relation, and solitons have been directly observed in experiments \cite{ds1,ds2,vs9,soliton4,vs10}. In single-component BECs, a bright soliton (BS) manifests as a localized atomic wave packet on a vanishing background and thus behaves as a quasiparticle with positive mass~\cite{bs1,bs2}. In contrast, a dark soliton is a localized density defect with a phase jump in a repulsive condensate, and its dynamics are characterized by a negative inertial mass~\cite{ds1,ds2,ds3,ds4}. In two-component BECs, the coupling between components enables the formation of a rich variety of vector solitons, such as degenerate and nondegenerate bright-bright solitons, dark-dark solitons, and combinations consisting of a bright soliton in one component and a dark soliton in the other \cite{vs5,vs6,vs8,vs11,vs12,bd}. The dynamics of bright-bright and dark-dark solitons resembles that of their scalar counterparts, exhibiting positive and negative inertial mass, respectively, while dark-bright solitons in attractive interactions are generally unstable. Interestingly, in nonintegrable systems, dark-bright (DB) solitons possess two distinct branches: one with positive mass and the other with negative mass, which can be mutually converted under external driving \cite{mlz2025,zhao2020,Jos2023}. However, in the Manakov model with repulsive interactions~\cite{manakov}, the DB solitons have long been regarded as having only negative inertial mass.

Recent studies report that even in the Manakov case with repulsive interactions, the DB solitons also support a positive mass branch, although both branches exhibit a dip in the total density~\cite{gx2025}. The positive mass branch solitons are broader and shallower than negative mass branch ones. Unlike the non-integrable case, these branches cannot be mutually converted under external driving. This results indicate that DB solitons generally have two inertial mass branches with different density and phase distributions, which is independent of integrability. This naturally raises the question: in an $N$-component BECs composed of one dark component and multiple bright components, how many branches do the vector soliton possess, and what are corresponding density and phase characteristics? When all bright components are degenerate, the system reduces to the two-component DB case, which exhibits only two branches-one with positive mass and the other with negative mass~\cite{dbb1,vs16}. Recent works reports nondegenerate vector solitons~\cite{vs3,vs31,vs4,ndbb,ns1,ns2,ns21,ns22,ns23}, where different components occupy distinct eigenstates of an effective quantum well~\cite{eff1,vs7}, lead us to expect that these nondegenerate solitons can support a richer branches.

In this work, we revisit nondegenerate dark-bright-bright solitons of the Manakov model with repulsive interactions, wherein the first component is a double-valley dark soliton and the second and third components are asymmetric bright solitons. We show that a nondegenerate dark-bright-bright soliton admits four distinct branches at the same velocity for fixed bright component particle numbers, two with positive effective mass and two with negative effective mass. The two positive negative mass pairs organize into two disjoint excitation energy velocity loops, with the lower energy nondegenerate solitons being typically broader and shallower. We determine that the maximum velocity depends only on the larger bright component particle number. A systematic stability analysis demonstrates that all four nondegenerate dark-bright-bright solitons branches are stable. In addition, numerical simulations of nondegenerate dark-bright-bright solitons under a constant external force reveal that these solitons can exhibit either positive or negative mass motion. However, once it is accelerated to the maximal velocity, the integrable system forbids conversion between positive and negative mass branches. The dark component splits into two dark solitons while the two bright components decouple, causing the nondegenerate soliton fission into two dark-bright solitons. We further extend this analysis to $N$-component Bose-Einstein condensates and show that nondegenerate vector solitons support $2^{N-1}$ inertial-mass branches, with $2^{N-2}$ branches of positive mass and $2^{N-2}$ branches of negative mass.  Each a positive mass branch and a negative mass branch form a closed dispersion relation loop, so that the nondegenerate vector soliton have $2^{N-2}$ disjoint loops.

The paper is organized as follows. In section \ref{soliton}, we incorporate the bright component particle numbers into the nondegenerate dark-bright-bright solitons for which four branches are obtained according to the relation between soliton's width and velocity. This analysis reveals four branches of solutions have different profiles, phase structures, and inertial masses. In section \ref{stability}, we examine the stability and dynamical behaviors under external driving. In section \ref{nbec}, we examine nondegenerate vector solitons in four-component Bose-Einstein condensates and subsequently generalize the analysis to $N$-component systems, revealing nondegenerate vector solitons have a varied inertial mass branches. Finally, in section \ref{conclusion}, we summarize and discuss the results.

\section{nondegenerate vector solitons with varied branches}\label{soliton}

\begin{figure}[t]
\begin{center}
\subfigure{\includegraphics[width=80mm]{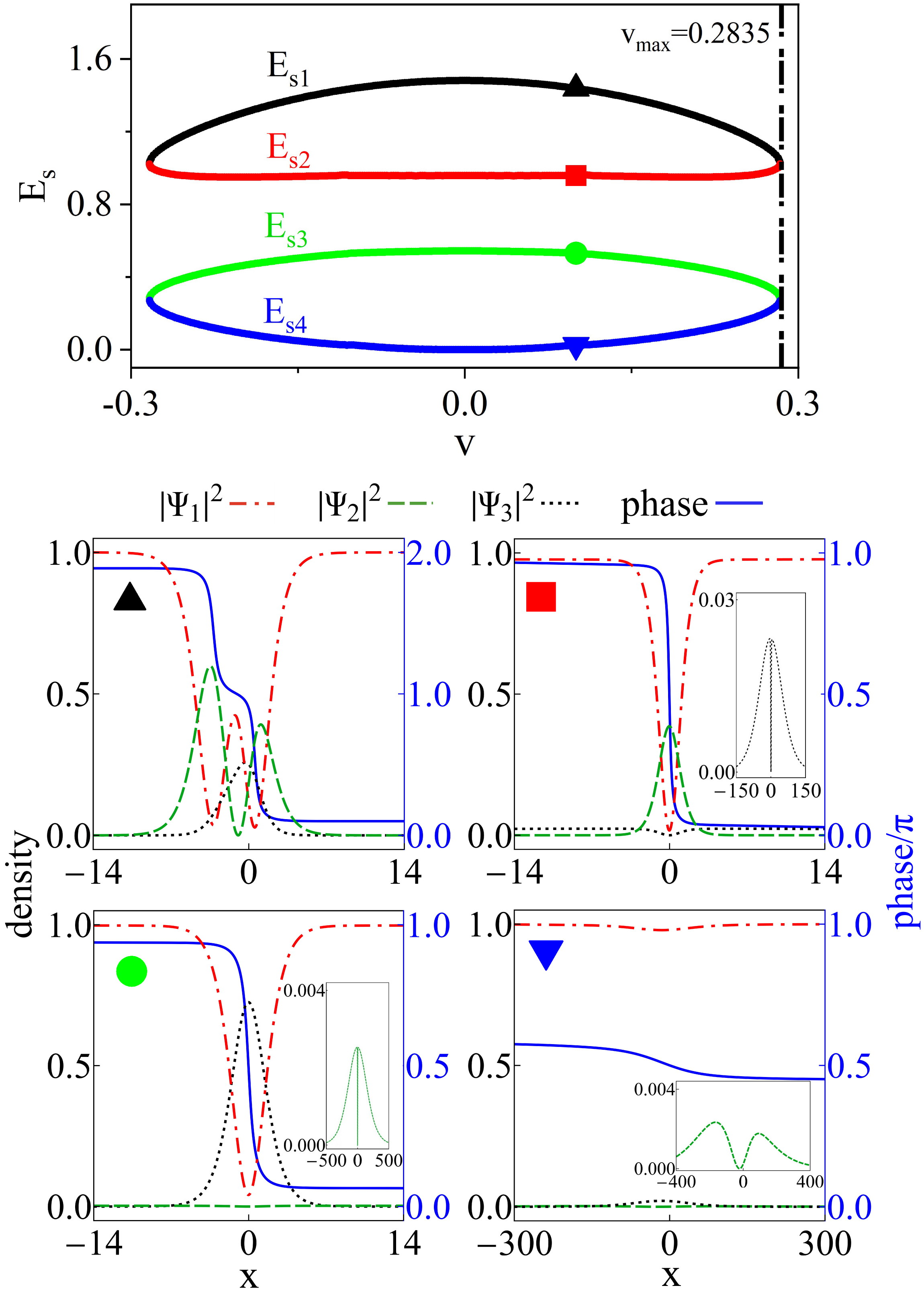}}
\end{center}
\caption{(Top) The energy-velocity dispersion relation of nondegenerate dark-bright-bright solitons with particle number for bright soliton $N_{B2}=1$ and $N_{B3}=3$. The nondegenerate soliton maximum velocity $v_{\max} = 0.2835$. The red and blue curves correspond to the positive mass branches, while the black and green curves correspond to the negative mass branches.
(Bottom) The density and phase distributions of the nondegenerate soliton with velocity $v=0.1$. The first component forms a double-valley dark soliton, while the second and third components are asymmetric bright solitons. Insets display the weakly localized tails on an expanded horizontal scale. Insets display the weakly localized solitons on an expanded horizontal scale.}\label{fig1}
\end{figure}

Under the mean-field approximation, the dynamics of a quasi-one-dimensional three-component BEC can be described by the following coupled Gross-Pitaevskii equations, whose dimensionless form can be written as~\cite{gp1,gp2,gp3}:
\begin{subequations}\label{CNLS}
\begin{align}
\nonumber i\frac{\partial \Psi_1}{\partial t}=&-\frac{1}{2}\frac{\partial^2\Psi_1}{\partial x^2}-V_1\Psi_1\\
&-(g_{11}|\Psi_1|^2+g_{12}|\Psi_2|^2+g_{13}|\Psi_3|^2)\Psi_1,\\
\nonumber i\frac{\partial \Psi_2}{\partial t}=&-\frac{1}{2}\frac{\partial^2\Psi_2}{\partial x^2}-V_2\Psi_2\\
&-(g_{12}|\Psi_1|^2+g_{22}|\Psi_2|^2+g_{23}|\Psi_3|^2)\Psi_2,\\
\nonumber i\frac{\partial \Psi_3}{\partial t}=&-\frac{1}{2}\frac{\partial^2\Psi_3}{\partial x^2}-V_3\Psi_3\\
&-(g_{13}|\Psi_1|^2+g_{23}|\Psi_2|^2+g_{33}|\Psi_3|^2)\Psi_3.
\end{align}
\end{subequations}
where $\Psi_i$ ($i=1,2,3$) represents the wave function of the $i$-th component, $x$ and $t$ respectively represent the spatial coordinate and time evolution. $V_i$ denotes the weak trapping potentials, which we set to zero $(V_i=0)$ in order to obtain analytic soliton solutions.
The coefficients $g_{ii}$ ($g_{ij}$) are the intra- (inter-) species interaction strengths given by the s-wave scattering length. If $g_{ii}$ and $g_{ij}$ are all equal, the equations will become the well-known Manakov model \cite{manakov}, which is completely integrable and can be solved exactly by the Darboux transformation \cite{darb1,darb2}, Hirota bilinear method \cite{hb}, and inverse-scattering method \cite{invs}, resulting in many different vector solitons \cite{vs1,vs2,vs3,vs4,vs5,vs6,vs7,vs8,vs11,vs12,vs14,vs15,vs16}. In two-component Bose-Einstein condensates, it has been established that the plane wave is modulationally unstable for attractive interactions $(g<0)$~\cite{mi}, so bright-dark (BD) solitons cannot maintain long term stability~\cite{bd}. Consequently, attention has focused on repulsive $(g>0)$ BECs, where dark-bright (DB) solitons exist~\cite{vs5,vs6,vs8,vs9,vs10,vs11,vs12}. Here the total density develops a dip, suggesting that the inertial mass should be negative.

Recent studies have demonstrated that DB solitons admit a positive mass branch whose dynamics differs from that of the negative mass branch~\cite{gx2025}. This naturally prompts the question, for three-component dark-bright-bright (DBB) solitons, whether positive-mass branches also exist and how many distinct branches arise. For the degenerate DBB solutions, the two bright components share a similar structure, reducing one degree of freedom; accordingly, the number of branches coincides with that of the two-component DB case. To determine the maximal branch count, we focus on the nondegenerate DBB soliton \cite{vs4}, which one component hosts a bright solitons with nodes, its soliton profiles differ from those of degenerate DBB solitons. The total density of nondegenerate DBB also exhibits dips. Those observation motivates our search for nondegenerate DBB solitons with positive inertial mass. The nondegenerate dark-bright-bright soliton can be written as (the background density is scaled to be 1):
\begin{subequations}\label{ndsoliton}
\begin{align}
\Psi_1&=\frac{N_{1}}{M} e^{-it},\\
\Psi_2&=-i 2 w_{1} \sqrt{1 - v^{2} - w_{1}^{2}}\frac{1}{\xi_{1}} \frac{N_{2}}{M}e^{\, i \left[ v x - \tfrac{1}{2} \left( 2 + v^{2} - w_{1}^{2} \right) t \right]},\\
\Psi_3&=-i 2 w_{2} \sqrt{1 - v^{2} - w_{2}^{2}}\frac{1}{\xi_{2}} \frac{N_{3}}{M}e^{\, i \left[ v x - \tfrac{1}{2} \left( 2 + v^{2} - w_{2}^{2} \right) t \right]}.
\end{align}
\end{subequations}
with
\begin{align}\label{ndsoliton1}
M &= \Big[e^{\kappa_{1}+\kappa_{2}}
+e^{\kappa_{1}-\kappa_{2}}
+e^{\kappa_{2}-\kappa_{1}} \Big] \gamma_1^2+ \gamma_2^2 e^{-\kappa_{1}-\kappa_{2}}, \notag\\[8pt]
N_{1} &= \Bigg[
\frac{\xi_{2}^{*}}{\xi_{2}}e^{\kappa_{1}-\kappa_{2}}
+ \frac{\xi_{1}^{*}}{\xi_{1}}e^{\kappa_{2}-\kappa_{1}} +e^{\kappa_{1}+\kappa_{2}} \notag \\[8pt]
&+\frac{\gamma_2^2\xi_{1}^{*}\xi_{2}^{*}}{4\xi_{1}\xi_{2}}e^{-\kappa_{1}-\kappa_{2}} \Bigg]\gamma_1^2, \notag \\[8pt]
N_{2} &= \gamma_1 \Big[
\gamma_1 e^{\kappa_{2}}
+ \gamma_2 e^{-\kappa_{2}} \Big],
N_{3} = \gamma_1 \Big[
\gamma_1 e^{\kappa_{1}}
+ \gamma_2 e^{-\kappa_{1}} \Big], \notag \\[4pt]
\kappa_{1} &= w_{1}(x-vt),\kappa_{2} = w_{2}(x-vt),\notag \\[4pt]
\xi_{1}&=iw_1-v,\xi_{2}=iw_2-v,
\gamma_{1}=w_1+w_2,\gamma_{2}=w_1-w_2.\notag
\end{align}
where $\Psi_1$ is double-valley dark soliton, $\Psi_2$ and $\Psi_3$ are asymmetric bright solitons. The parameters $w_{j}$ ($j=1,2$) are inverse widths (larger $w_{j}$ implies a narrower soliton), and $w_{1}\neq w_{2}$. We define the particle number of the $k$-th bright component as
$N_{Bk}\equiv\int_{-\infty}^{\infty}\!|\Psi_k(x)|^{2}\,dx$ $(k=2,3)$, which is related to the parameter set $(v, w_1, w_2)$. Those parameters and particle numbers are required to satisfy the following relations:
\begin{subequations}\label{ndsolitonrestriction}
\begin{align}
N_{B2}=\frac{2(w_1-v^2w_1-w_1^3)}{v^2 + w_1^2},\\
N_{B3}=\frac{2(w_2-v^2w_2-w_2^3)}{v^2 + w_2^2}.
\end{align}
\end{subequations}

\begin{figure}[t]
\begin{center}
\subfigure{\includegraphics[width=87mm]{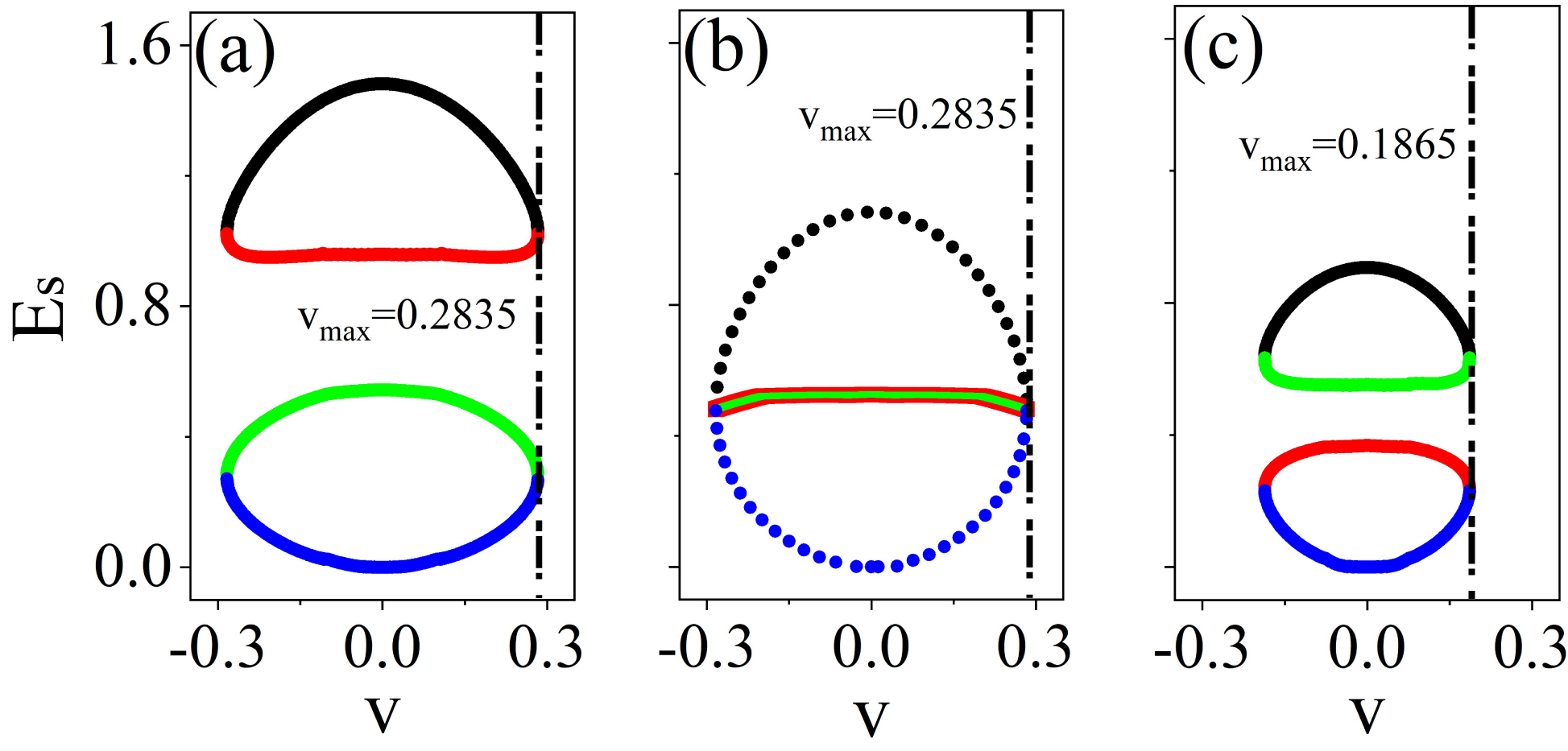}}
\end{center}
\caption{The dispersion relations $E_s(v)$ of nondegenerate dark-bright-bright solitons for varying $N_{B2}$.
The black, red, green, and blue correspond to the branches $E_{s1}$, $E_{s2}$, $E_{s3}$, and $E_{s4}$, respectively.
(a) Case $N_{B2}<N_{B3}$: red and blue denote positive inertial mass branches, whereas black and green denote negative inertial mass branches. The common maximal velocity is $v_{\max}=0.2835$.
(b) Case $N_{B2}=N_{B3}$: the red and green curves coincide, indicating energy degeneracy of $E_{s2}$ and $E_{s3}$. The dotted curves correspond to the limiting dispersions of $E_{s1}$ and $E_{s4}$ branches as $N_{B2}=N_{B3}+10^{-5}$. The maximal velocity remains $v_{\max}=0.2835$.
(c) Case $N_{B2}>N_{B3}$: green and blue correspond to positive inertial mass branches, while black and red correspond to negative inertial mass branches. Here $v_{\max}=0.1865$.
Other parameters: (a) $N_{B2}=1$; (b) $N_{B2}=3$; (c) $N_{B2}=5$ (with the stated $N_{B3}=3$ in each panel).}\label{fig2}
\end{figure}

We choose the $N_{B2}$, $N_{B3}$ and $v$ to be independent parameters, considering that they are much easier to be measured in experiments.
Eqs.~\eqref{ndsolitonrestriction} yield two independent cubic equations for the inverse widths $w_1$ and $w_2$; thus, a given $(N_{B2},N_{B3},v)$ generally produces three roots for each $w_j$ $(j=1,2)$. Among the three roots for $w_j$, one is negative in each case and violates the condition for nondegenerate solitons, $|w_j|>1$.
Accordingly, although each cubic admits three mathematical roots, only two values of $w_1$ and two values of $w_2$ are physically admissible. Any combination of $w_1$ and $w_2$ is a valid soliton solution of the system. Consequently, for any prescribed $(N_{B2}, N_{B3}, v)$, there exist four physically admissible pairs $(w_1, w_2)$, corresponding to four soliton solutions with distinct widths.
\begin{figure}[t]
\begin{center}
\subfigure{\includegraphics[width=85mm]{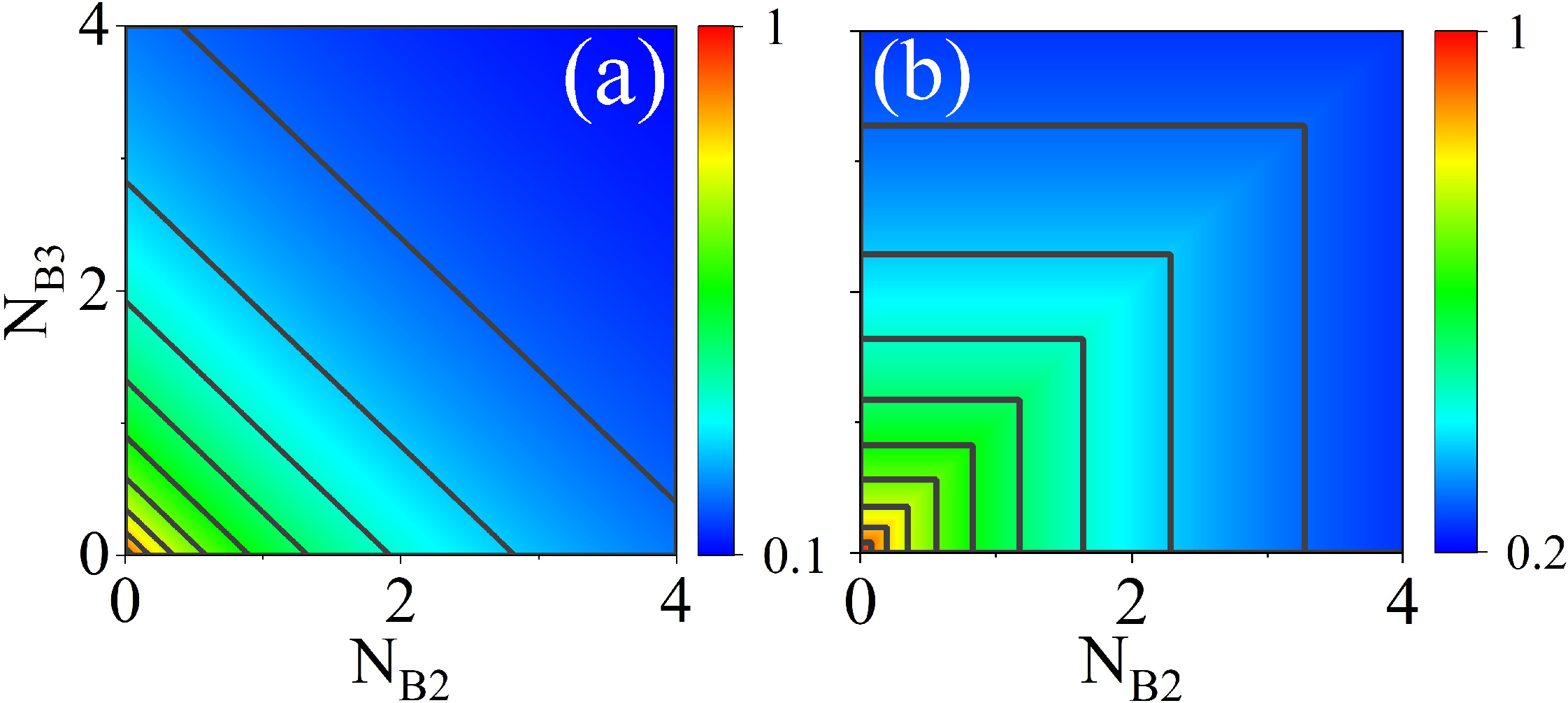}}
\end{center}
\caption{The maximum velocity of degenerate (a) and nondegenerate (b) dark-bright-bright solitons with varying $N_{B}$ in the region [0,4]. The black solid curves denote iso-velocity lines. For degenerate solitons, the maximum velocity is governed by the total bright component particle number, whereas for nondegenerate solitons it is controlled by the larger bright component particle number.
}\label{fig3}
\end{figure}

A soliton can regarded as a quasiparticle whose dynamical behavior is dictated by its dispersion relation; hence, computing the dispersion relations of the four branches solitons is essential. The excitation energy of soliton can be given by $E_s=\int \frac{1}{2}|\partial_x\Psi_1|^2+\frac{1}{2}|\partial_x\Psi_2|^2+\frac{1}{2}|\partial_x\Psi_3|^2+\frac{1}{2}|\Psi_1|^4+\frac{1}{2}|\Psi_2|^4+\frac{1}{2}|\Psi_3|^4+(|\Psi_1|^2-1)|\Psi_2|^2+(|\Psi_1|^2-1)|\Psi_3|^2+|\Psi_2|^2|\Psi_3|^2dx$. The four soliton solutions with distinct widths give four excitation energies, in descending order, $E_{s1}$, $E_{s2}$, $E_{s3}$, and $E_{s4}$. As a example, Fig.~\ref{fig1} (Top) shows the excitation energy-velocity dispersion $E_s(v)$ for a nondegenerate soliton, where the black, red, green and blue curve denote the $E_{s1}$, $E_{s2}$, $E_{s3}$, and $E_{s4}$ branches, respectively. The four branches share the same maximum velocity, which energies become pairwise degenerate at this maximum velocity. The nondegenerate soliton dispersion consists of two nonintersecting loops, in contrast to the single loop characteristic of the degenerate solitons \cite{gx2025,mlz2025,zhao2020,Jos2023}. As a quasiparticle excitation, the soliton's inertial mass is a key characteristic of its dynamical behavior. The inertial mass can be derived as $M^*= 2\partial Es/\partial (v^2) $ \cite{es1,es2}, the $E_{s1}$ and $E_{s3}$ branches have negative inertial mass, whereas the $E_{s2}$ and $E_{s4}$ branches have positive inertial mass.

Because the inverse widths $w_j$ differ across branches, the four solutions at the same velocity possess distinct soliton profiles, previous studies examined only the $E_{s1}$ branch \cite{vs3,vs4}. Fig.~\ref{fig1} (Bottom) displays the density profiles and phase distributions of the four branches at $v=0.1$. The first component forms an double-valley dark soliton, whereas the second and third components exhibit asymmetric bright solitons. Relative to $E_{s1}$ branch, the double-hump solitons components in $E_{s2}$ and $E_{s3}$ branches are already broader and shallower; the lowest energy branch $E_{s4}$ having the widest and most weakly localized profiles across all four components. The four branches solitons admit different phase distributions of dark solitons. In particular, the width of double hump bright tends to be infinite for the $E_{s2}$ and $E_{s3}$ branches with zero moving velocity. Because the particle numbers held fixed, the corresponding soliton amplitudes become very small, whereby the nondegenerate solution approaches to the dark-bright soliton solution.

Since the exact solutions impose no restriction on the particle number $N_{B}$, and contemporary experimental techniques allow precise loading of particles into the components, we further analyze the soliton dispersion relations for various $N_{B}$ settings based on Eqs.~\eqref{ndsolitonrestriction}. In the limit $\max\{N_{B2},N_{B3}\}\!\to\!0$, the bright soliton in the second and third components vanish, consequently, the dispersion relation tends to be that of a scalar dark soliton. The energy gap between the two loops is governed by $|N_{B2}-N_{B3}|$: as the two bright component particle numbers approach each other, the energy gap between the two loops decreases. Especially, at the case of $N_{B2}=N_{B3}$, the branches of $E_{s2}$ and $E_{s3}$ become energy degenerate, and the solitons inertial mass is very small, as shown in Fig.~\ref{fig2} (b). In this case, the two solutions differ only by swapping the bright component wave functions: specifically, the second component of $E_{s2}$ is the same double-hump bright soliton as the third component of $E_{s3}$, and the third component of $E_{s2}$ is the same asymmetric bright soliton as the second component of $E_{s3}$. Because the Manakov system have identical intra- and inter-component couplings, this exchange of the two bright soliton components does not alter any property of the solution. The other branches $E_{s1}$ and $E_{s4}$ would require $w_1=w_2$ at this point, which violates the nondegenerate dark-bright-bright solitons condition. The dotted curve in Fig.~\ref{fig2} (b) indicates the limiting dispersion relation for \(N_{B2}\rightarrow N_{B3}\). Upon further increasing the particle number of the second bright component $N_{B2}$, and the condition $N_{B2}>N_{B3}$ is satisfied, one finds $E_{s2}<E_{s3}$. The branches of $E_{s1}$ and $E_{s3}$ recombine a new loop, whereas $E_{s2}$ and $E_{s4}$ constitute the other, as shown in Fig.~\ref{fig2}(c). Therefore the $E_{s2}$ branch becomes the negative mass branch while $E_{s3}$ becomes the positive mass branch. The dispersion relation (loop size) shrinks monotonically with $N_{B2}$ increases.

For scalar dark solitons, the maximal velocity to be no greater than the condensate sound velocity. For DB solitons, the presence of the bright component reduces the maximal velocity below the sound velocity, with its value set by the bright component particle number. In three-component BECs, both degenerate and nondegenerate DBB solitons exist. Our next objective is to determine whether the maximal velocity differ at fixed particle numbers and to quantify how these maximal velocity depend on the bright component particle numbers. Because the velocity limit is not directly readable from its exact solution, we determine the maximum velocity $v_{\max}$ by simultaneously solving \eqref{ndsolitonrestriction} under the requirement that the inverse widths remain physically meaningful. When $v>v_{\max}$, the inverse widths become a complex number. We find $v_{\max}$ to be lower than that of a scalar dark soliton \cite{ds1,ds2,ds3,ds4}. Fig.~\ref{fig3} displays $v_{\max}$ for both degenerate and nondegenerate solitons under different $(N_{B2},N_{B3})$ settings. For degenerate dark-bright-bright solitons, $v_{\max}$ is jointly controlled by $N_{B2}$ and $N_{B3}$ and decreases as the total particle number $N_{B2}+N_{B3}$ increases. By contrast, for nondegenerate solitons, $v_{\max}$ is governed by $\max\{N_{B2},N_{B3}\}$. The underlying mechanism for the maximum velocity still needs further study.

\section{stability and acceleration motion of nondegenerate vector solitons}\label{stability}

\begin{figure}[t]
\begin{center}
\subfigure{\includegraphics[width=87mm]{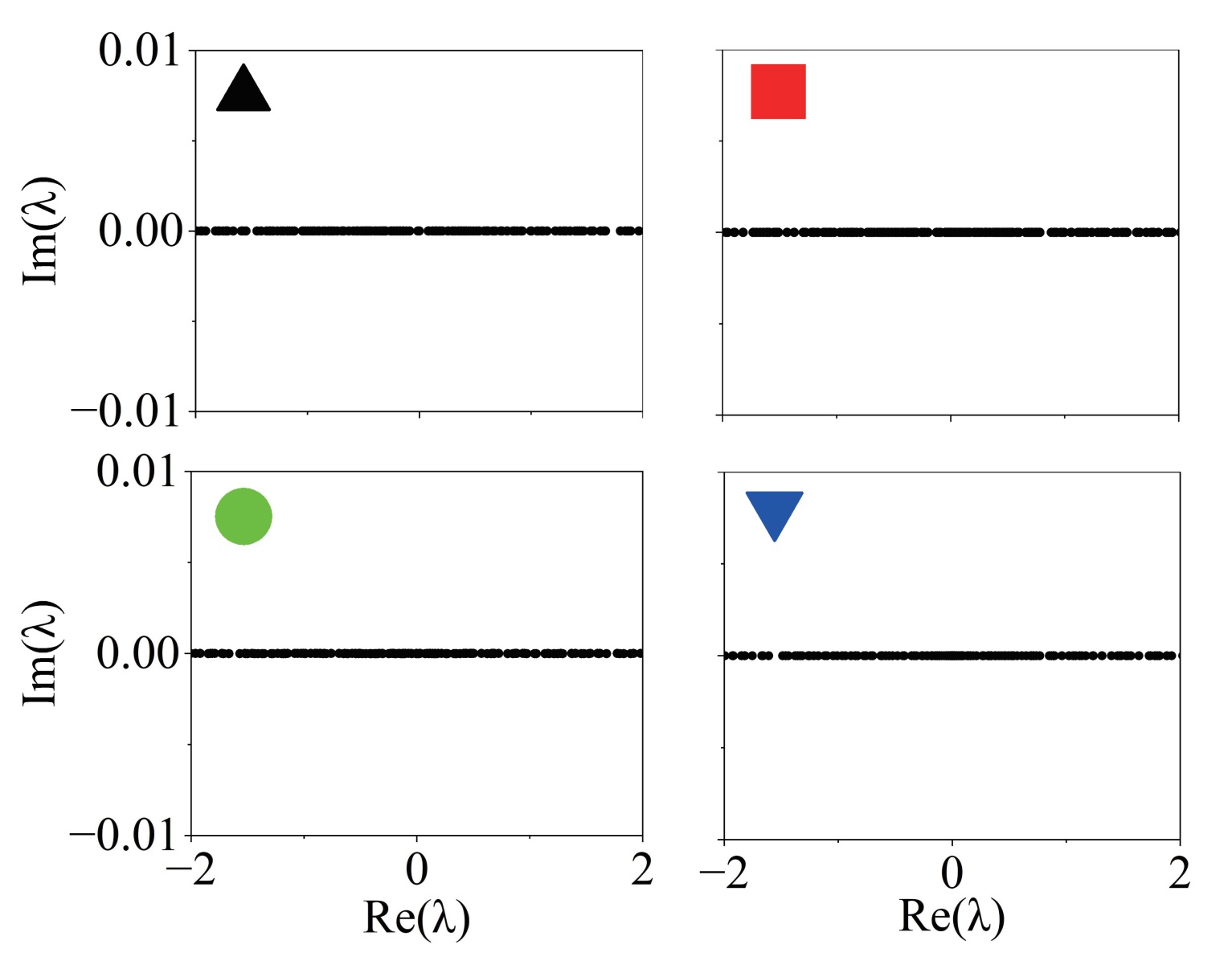}}
\end{center}
\caption{Bogoliubov-de Gennes excitation spectra of four nondegenerate dark-bright-bright solitons with $v=0.1$. They exhibit a spectrum without imaginary part. The parameters are the same as symbols in Fig.~(\ref{fig1}).}\label{fig4}
\end{figure}

We need to check stability of nondegenerate vector solitons with four branches, since they were not observed in neither previous theoretical nor experimental studies. The Bogoliubov-de Gennes excitation spectrum is one of the commonly used methods for testing the stability of solitons. Since the solitons could admit a non-zero velocity, we introduce the comoving coordinate $\tilde{x}=x-vt$ (with $v$ the soliton velocity). The nondegenerate vector soliton solution becomes $\Psi_{1c}=\phi_1(\tilde{x})e^{iv\tilde{x}}e^{-i\mu t}$, $\Psi_{2c}=\phi_2(\tilde{x})e^{-i\gamma_1 t}$ and $\Psi_{3c}=\phi_3(\tilde{x})e^{-i\gamma_2 t}$, where $\phi_1(\tilde{x})=N_{1}/M$, $\phi_2(\tilde{x})=-i2w_{1}\sqrt{1-v^{2}-w_{1}^{2}}N_{2}/(\xi_{1}M)$, $\phi_3(\tilde{x})=-i2w_{2}\sqrt{1-v^{2}-w_{1}^{2}}N_{3}/(\xi_{2}M)$, $\kappa_{1}=w_{1}\tilde{x}$ and $\kappa_{2}=w_{2}\tilde{x}$. We can consider the weakly perturbed solutions of Eq.~(\ref{CNLS}):

\begin{subequations}\label{perturbed}
\begin{align}
\Psi_{1p}&=[\phi_1(\tilde{x})+\epsilon_1(\tilde{x},t)]e^{iv\tilde{x}}e^{-i\mu t},\\
\Psi_{2p}&=[\phi_2(\tilde{x})+\epsilon_2(\tilde{x},t)]e^{-i\gamma_1 t},\\
\Psi_{3p}&=[\phi_3(\tilde{x})+\epsilon_3(\tilde{x},t)]e^{-i\gamma_2 t}.
\end{align}
\end{subequations}
where $\mu=1+v^2/2$, $\gamma_1=1-w_1^2/2$ and $\gamma_2=1-w_2^2/2$. $\epsilon$ denote small perturbations:
\begin{subequations}\label{perturbations}
\begin{align}
\epsilon_1(\tilde{x},t)=P_1(\tilde{x})e^{i\lambda t}+P_2^*(\tilde{x})e^{-i\lambda^* t},\\
\epsilon_2(\tilde{x},t)=Q_1(\tilde{x})e^{i\lambda t}+Q_2^*(\tilde{x})e^{-i\lambda^* t},\\
\epsilon_3(\tilde{x},t)=L_1(\tilde{x})e^{i\lambda t}+L_2^*(\tilde{x})e^{-i\lambda^* t}.
\end{align}
\end{subequations}
where $\lambda=\lambda_R+i\lambda_I$ is the eigenvalue, and $(P_1,P_2,Q_1,Q_2,L_1,L_2)^T$ is the eigenvector. By substituting weakly perturbed solutions into Eq.~(\ref{CNLS}) and ignoring the high-order terms of the perturbation, we obtain an eigenvalue problem for the eigenvector. Generally, if the eigenvalue $\lambda$ is purely real, it indicates that the soliton is stable against small perturbations. Otherwise, if the eigenvalue of $\lambda$ has non-zero imaginary part, it indicates that the soliton is unstable and will be destroyed by small perturbations that grow exponentially. This spectral problem can be solved numerically, the results are shown in Fig.~(\ref{fig4}). One can find that the four branches of nondegenerate soliton solutions only exist spectra without imaginary part, so those branches are stability and can be stable over a long time evolution. Since all four branches are spectrally stable, their experimental realization feasible. Moreover, one can systematic studies of these solitons dynamics under a variety of external potentials.

\begin{figure}[t]
\begin{center}
\subfigure{\includegraphics[width=88mm]{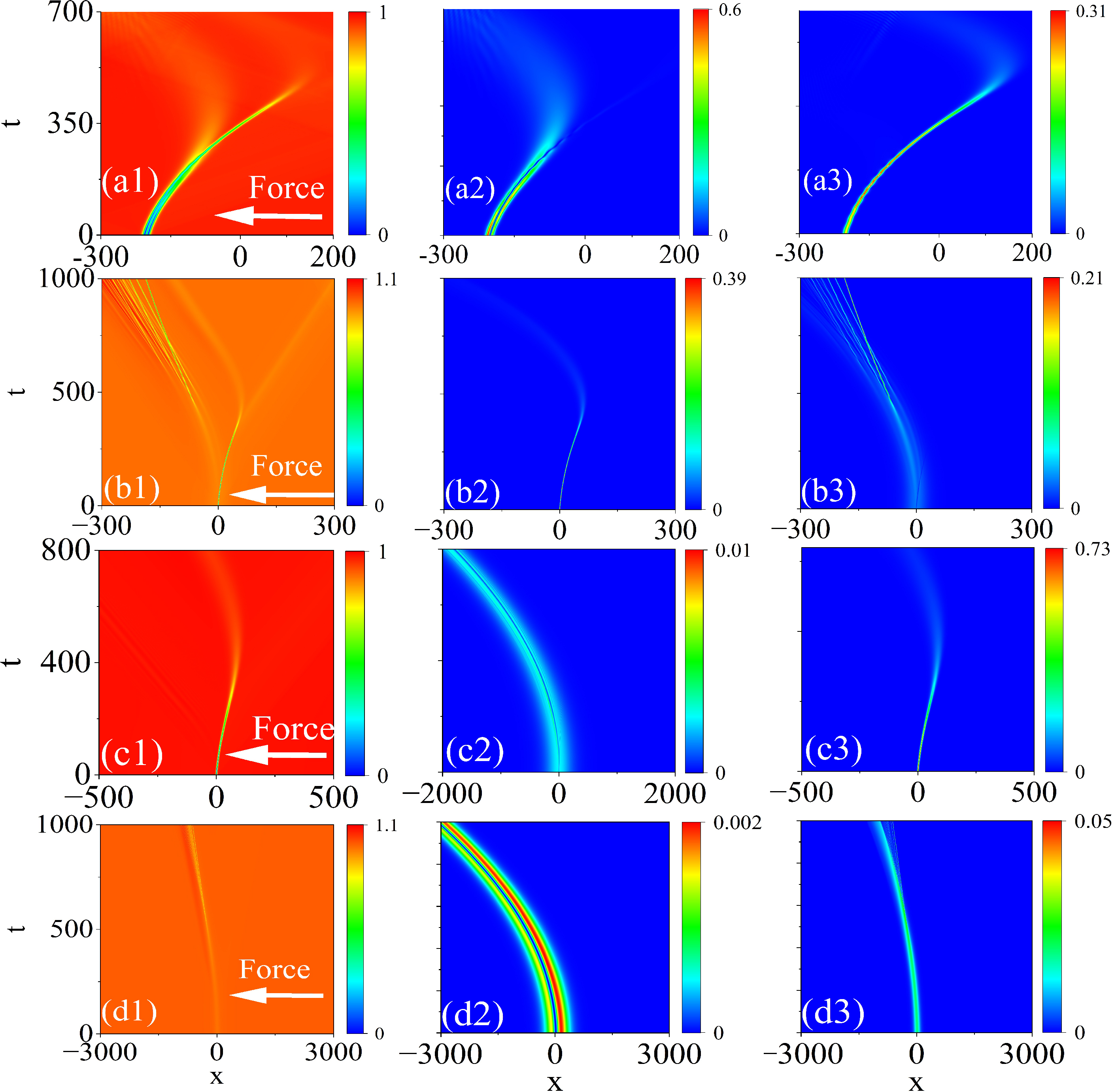}}
\end{center}
\caption{Density evolutions of nondegenerate dark-bright-bright solitons with different branches driven by $F_2=0.003$ and $F_3=0.001$. (a1)-(a3) The nondegenerate soliton of $E_{s1}$ with negative mass is accelerated and diffuses two negative mass dark-bright solitons.
(b1)-(b3) The nondegenerate soliton of $E_{s2}$ with positive mass is accelerated and diffuses a negative and a positive mass dark-bright solitons. (c1)-(c3) The nondegenerate soliton of $E_{s3}$ with negative mass is accelerated and diffuses a negative and a positive mass dark-bright solitons. (d1)-(d3) The nondegenerate soliton of $E_{s4}$ with positive mass is accelerated and diffuses two positive mass dark-bright solitons. The parameters are the same as the symbols in the Fig.~(\ref{fig1}), respectively.}\label{fig5}
\end{figure}

The dynamics of vector solitons under external potentials have been extensively investigated. In harmonic trapping potentials, vector solitons exhibit oscillatory motion~\cite{vs9,bdtrap2,bdtrap3}, whereas in linear (constant gradient) external potentials, oscillations emerge only in nonintegrable systems~\cite{Jos2023,mlz2025,gx2025,zhao2020}. However, a transition between these branches is prohibited, and no oscillation occurs in the same potential in integrable system. As the soliton approaches its maximum velocity, negative mass solitons undergo diffusion, whereas positive mass solitons undergo splitting \cite{gx2025}. The above nondegenerate solitons also have the two positive mass branches and the two negative mass branches, these observations motivate us to examine the dynamics of nondegenerate dark-bright-bright solitons under linear external potentials. We apply a constant weak force on the bright soliton by adding a linear external potential $V_2(x)=-F_2x$ and $V_3(x)=-F_3x$ on the bright soliton component. The initial states of four solitons are given by Eq.~(\ref{ndsoliton}) with $v=0.1$, $N_{B2}=1$ and $N_{B3}=3$. The numerical simulation results for the dynamical evolution of these four types of nondegenerate solitons are shown in Fig.~\ref{fig5} with a weak external force $F_2 = 0.003$ and $F_3 = 0.001$ (the force is along the negative $x$-axis direction), respectively. Among the four branches, the $E_{s1}$ and $E_{s4}$ branches soliton evolve along their respective excitation energy dispersion curves under a weak linear potential. Since the $E_{s1}$ branch soliton carries a negative effective mass, the soliton undergoes accelerated motion toward the positive $x$-direction. As its velocity approaches the maximal value, the transition between negative- and positive-mass branches is forbidden in integrable systems; consequently, no oscillatory motion occurs, in contrast to the non-integrable case~\cite{Jos2023,mlz2025}. At this stage, the dark component undergoes split, while the two bright components decouple and form two negative mass dark-bright solitons. In contrast, the $E_{s4}$ branch possesses a positive effective mass and therefore decelerates toward the negative $x$-direction. When the velocity decreases to zero, the nondegenerate DBB soliton transforms into two positive mass dark-bright solitons. For the intermediate branches $E_{s2}$ and $E_{s3}$, the application of the external potential induces a rapid conversion into one positive mass and one negative mass dark-bright soliton. The subsequent dynamics of these dark-bright solitons follow the behavior of two-component DB solitons~\cite{gx2025}. For the dark-bright soliton of negative mass, it will diffuse after reaching its maximum velocity; for the dark-bright soliton of positive mass, the soliton splits into several small soliton fragments near the maximum velocity.

\section{varied branches of nondegenerate vector solitons in $N$-components BECs}\label{nbec}
Previous studies have shown that dark bright solitons in two-component BECs admit two solution branches \cite{gx2025}. In the preceding section, we demonstrated that nondegenerate dark-bright-bright solitons in three-component BECs admit four branches. These observations motivate a systematic inquiry into the number of branches for nondegenerate solitons in $N$-component BECs. We frist analyze the four-component case. For the four-component Gross-Pitaevskii system, the nondegenerate dark-bright-bright-bright soliton can be written as (the background density is scaled to be 1):
\begin{small}\begin{subequations}\label{ndsoliton}
\begin{align}
\Psi_1&=\frac{J_{1}}{K} e^{-it},\\
\Psi_2&=-i 2 w_{1} \sqrt{1 - v^{2} - w_{1}^{2}}\frac{1}{\xi_{1}} \frac{J_{2}}{K}e^{\, i \left[ v x - \tfrac{1}{2} \left( 2 + v^{2} - w_{1}^{2} \right) t \right]},\\
\Psi_3&=-i 2 w_{2} \sqrt{1 - v^{2} - w_{2}^{2}}\frac{1}{\xi_{2}} \frac{J_{3}}{K}e^{\, i \left[ v x - \tfrac{1}{2} \left( 2 + v^{2} - w_{2}^{2} \right) t \right]},\\
\Psi_4&=-i 2 w_{3} \sqrt{1 - v^{2} - w_{3}^{2}}\frac{1}{\xi_{3}} \frac{J_{4}}{K}e^{\, i \left[ v x - \tfrac{1}{2} \left( 2 + v^{2} - w_{3}^{2} \right) t \right]},
\end{align}
\end{subequations}\end{small}
with
\begin{align}\label{ndsoliton1}
K &= \gamma_2^{2}\gamma_4^{2}\gamma_6^{2}
+ \gamma_1^{2}\gamma_6^{2}\gamma_3^{2}\, e^{2\alpha_1}
+ \gamma_1^{2}\gamma_4^{2}\gamma_5^{2}\, e^{2\alpha_2} \notag\\
&+ \gamma_2^{2}\gamma_3^{2}\gamma_5^{2}\, e^{2\alpha_3}+ \gamma_1^{2}\gamma_3^{2}\gamma_5^{2}[e^{2(\alpha_1+\alpha_2)} \notag\\
&+ e^{2(\alpha_1+\alpha_3)}+ e^{2(\alpha_2+\alpha_3)}
+ e^{2(\alpha_1+\alpha_2+\alpha_3)}], \notag\\
J_{1} &=
\frac{1}{(v-\mathrm{i}w_{1})(v-\mathrm{i}w_{2})(v-\mathrm{i}w_{3})} \notag\\
&\Big[
(v+\mathrm{i}w_{1})(v+\mathrm{i}w_{2})(v+\mathrm{i}w_{3})\,\gamma_{2}^{2}\gamma_{4}^{2}\gamma_{6}^{2} \notag\\
&+ (v-\mathrm{i}w_{1})(v+\mathrm{i}w_{2})(v+\mathrm{i}w_{3})\,\gamma_{1}^{2}\gamma_{6}^{2}\gamma_{3}^{2}\, e^{2\alpha_{1}} \notag\\
&+ (v+\mathrm{i}w_{1})(v-\mathrm{i}w_{2})(v+\mathrm{i}w_{3})\,\gamma_{1}^{2}\gamma_{4}^{2}\gamma_{5}^{2}\, e^{2\alpha_{2}} \notag\\
&+ (v+\mathrm{i}w_{1})(v+\mathrm{i}w_{2})(v-\mathrm{i}w_{3})\,\gamma_{2}^{2}\gamma_{3}^{2}\gamma_{5}^{2}\, e^{2\alpha_{3}} \notag\\
&+ \gamma_{1}^{2}\gamma_{3}^{2}\gamma_{5}^{2}\Big(
 (v-\mathrm{i}w_{1})(v-\mathrm{i}w_{2})(v+\mathrm{i}w_{3})\, e^{2(\alpha_{1}+\alpha_{2})} \notag\\
&+ (v-\mathrm{i}w_{1})(v+\mathrm{i}w_{2})(v-\mathrm{i}w_{3})\, e^{2(\alpha_{1}+\alpha_{3})} \notag\\
&+ (v+\mathrm{i}w_{1})(v-\mathrm{i}w_{2})(v-\mathrm{i}w_{3})\, e^{2(\alpha_{2}+\alpha_{3})} \notag\\
&+ (v-\mathrm{i}w_{1})(v-\mathrm{i}w_{2})(v-\mathrm{i}w_{3})\, e^{2(\alpha_{1}+\alpha_{2}+\alpha_{3})}
\Big)
\Big],\notag \\[8pt]
J_{2} &= e^{\alpha_{1}}\,
\gamma_{1}\gamma_{3}\Big[
\gamma_{2}\gamma_{4}\,\gamma_{6}^{\,2}
\;+\;
\gamma_{5}^{\,2}\big(
\gamma_{1}\gamma_{4}\,e^{2\alpha_{2}}
+\gamma_{2}\gamma_{3}\,e^{2\alpha_{3}} \notag\\
&+\gamma_{1}\gamma_{3}\,e^{2(\alpha_{2}+\alpha_{3})}
\big)
\Big],\notag\\[4pt]
J_{3} &= e^{\alpha_{2}}\gamma_{1}\Big\{
\gamma_{5}\gamma_{6}\big[-\gamma_{2}\gamma_{4}^{2}
+\gamma_{1}\gamma_{3}^{2}e^{2\alpha_{1}}\big]
+\gamma_{3}^{2}\gamma_{5}^{2}\big[-\gamma_{2}e^{2\alpha_{3}}\notag\\
&+\gamma_{1}e^{2(\alpha_{1}+\alpha_{3})}\big]
\Big\},\notag \\[4pt]
J_{4} &= e^{\alpha_{3}}\,
\gamma_{3}\gamma_{5}\Big[
\gamma_{2}^{2}\gamma_{4}\gamma_{6}
-\gamma_{1}^{2}\gamma_{3}\gamma_{6}\,e^{2\alpha_{1}}\,\alpha^{2}
-\gamma_{1}^{2}\gamma_{4}\gamma_{5}\,e^{2\alpha_{2}}\,\beta^{2}\notag\\
&+\gamma_{1}^{2}\gamma_{3}\gamma_{5}\,e^{2(\alpha_{1}+\alpha_{2})}\,\alpha^{2}\beta^{2}
\Big], \notag \\[4pt]
\xi_{1}&=iw_1-v,\xi_{2}=iw_2-v,
\gamma_{1}=w_1+w_2,\gamma_{2}=w_1-w_2,\notag \\[4pt]
\gamma_{3}&=w_1+w_3,\gamma_{4}=w_1-w_3, \gamma_{5}=w_2+w_3,\gamma_{6}=w_2-w_3,\notag\\
\alpha_{1} &= w_{1}(x-vt),\alpha_{2} = w_{2}(x-vt),\alpha_{3} = w_{3}(x-vt).\notag
\end{align}
where $\Psi_1$ is triple-valley dark soliton, $\Psi_2$, $\Psi_3$ and $\Psi_4$ are asymmetric bright solitons, $w_j>0$ ($j=1,2,3$) are soliton inverse widths related parameters, and $w_1\neq w_2\neq w_3$. We define the particle number of the $k$-th bright component$(k=2,3,4)$, which is determined by the parameter set $(v, w_1, w_2, w_3)$. Those parameters and particle numbers are required to satisfy the following relations:
\begin{subequations}\label{4ndsolitonrestriction}
\begin{align}
N_{B2}=\frac{2(w_1-v^2w_1-w_1^3)}{v^2 + w_1^2},\\
N_{B3}=\frac{2(w_2-v^2w_2-w_2^3)}{v^2 + w_2^2},\\
N_{B4}=\frac{2(w_3-v^2w_3-w_3^3)}{v^2 + w_3^2}.
\end{align}
\end{subequations}
As before, we take the bright component particle numbers $N_{B2}$, $N_{B3}$, $N_{B4}$ together with the soliton velocity $v$ as the independent parameters, since they are directly accessible in experiments. Eqs. \eqref{4ndsolitonrestriction} give three cubic constraints for the inverse widths $w_1$, $w_2$, and $w_3$, implying that a given $(N_{B2},N_{B3},N_{B4},v)$ generally yields three algebraic roots for each $w_j$ $(j=1,2,3)$. Although each cubic equation admits three mathematical roots, only two are physically admissible. Accordingly, each constraint gives two acceptable values of $w_j$, and any combination $(w_1,w_2,w_3)$ drawn from these admissible values constitutes a valid soliton solution. Consequently, for any $(N_{B2},N_{B3},N_{B4},v)$, there exist $2^3=8$ physically admissible triplets $(w_1,w_2,w_3)$, corresponding to eight soliton solutions with distinct inverse widths.

\begin{figure}[t]
\begin{center}
\subfigure{\includegraphics[width=75mm]{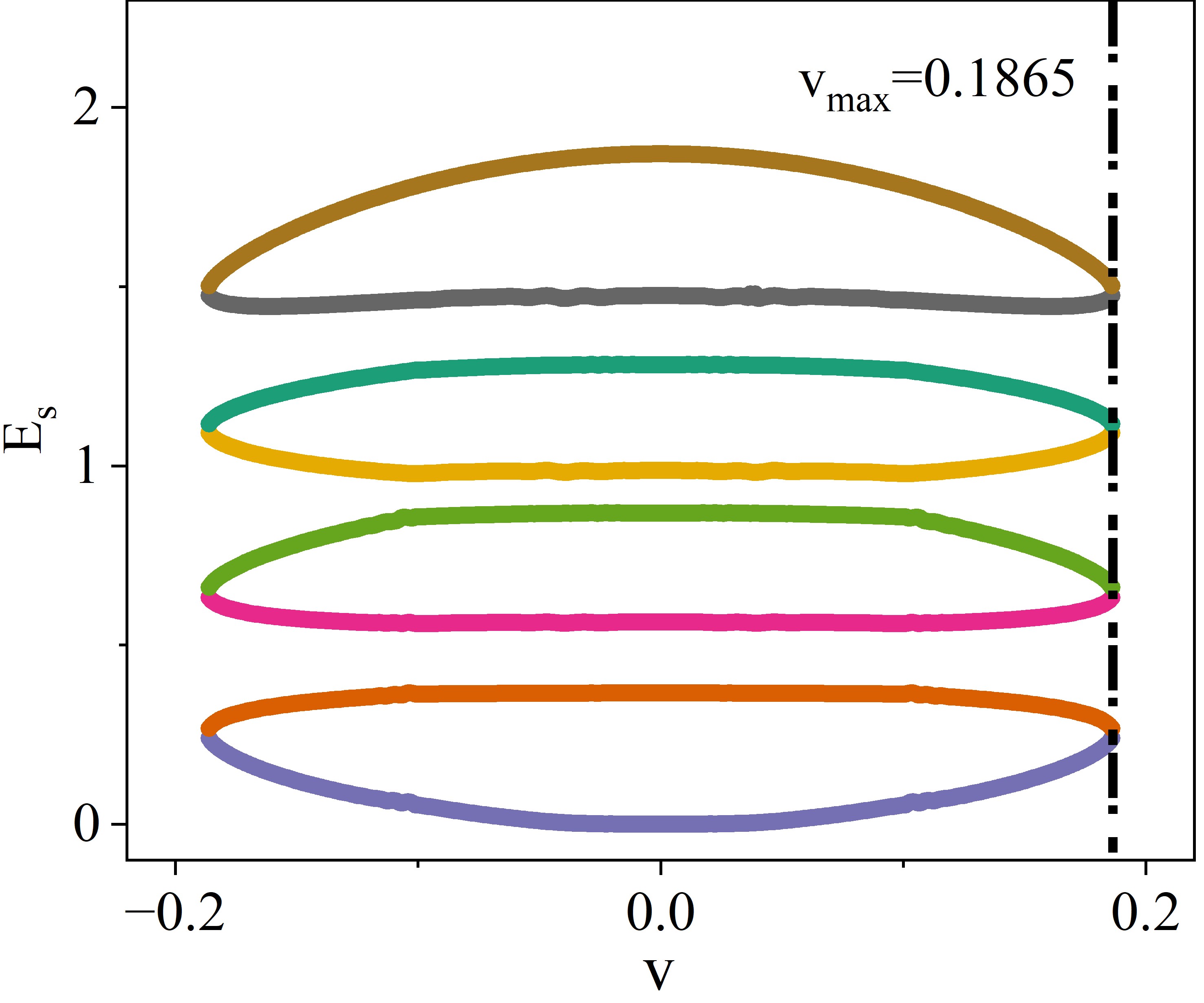}}
\end{center}
\caption{The energy-velocity dispersion relation of the nondegenerate dark-bright-bright-bright soliton with particle number for $N_{B1}=1$,$N_{B2}=3$ and $N_{B3}=5$, here the maximum velocity $v_{max}$=0.1865, determined by the largest bright component particle number. Curves of different colors denote distinct width branches of the soliton. For every loop in the dispersion, the higher energy branch is the negative inertial mass solitons, while the lower energy branch corresponds to positive inertial mass soliton.}\label{fig6}
\end{figure}

We analyze the excitation energies of the eight nondegenerate solutions and find that they generate eight distinct energy branches. As an illustration, Fig.~\ref{fig6} displays the excitation energy-velocity dispersion $ E_s(v)$ for the nondegenerate families, whereas earlier studies focused on the highest energy branch. All eight branches share the same maximum velocity, set by the largest bright component particle number, at which the energies become pairwise degenerate; accordingly, the nondegenerate solitons dispersion consists of four nonintersecting loops. Among these branches, four possess positive inertial mass and the remaining four possess negative inertial mass.

Next, we analyze the nondegenerate soliton sector of the $N$-component Gross-Pitaevskii equations. A nondegenerate soliton comprises $N-1$ bright components (indexed by $j=2,\ldots,N$) coupled to a single dark component \cite{wbh2024}. The particle number $N_{Bj}$ of the $j$-th bright component must satisfy:
\begin{eqnarray}\label{Nrestriction}
N_{Bj}=\frac{2\bigl(w_{j-1} - v^{2} w_{j-1} - w_{j-1}^{3}\bigr)}{v^{2}+w_{j-1}^{2}},
\end{eqnarray}
where $j=2,3,4,\ldots,N$, each equation yields three mathematical roots for $w_{j-1}$, of which two are physically admissible. Hence every bright component contributes two admissible inverse widths, and the total number of branches for an $N$-component nondegenerate soliton is:
\begin{eqnarray}\label{MNrestriction}
C_{N}=2^{\,N-1}.
\end{eqnarray}
These solutions exhibit mutually distinct density profiles and phase structures, underscoring the richness of nondegenerate soliton branches. For the $N$-component nondegenerate soliton, the excitation energy velocity dispersion organizes into $2^{N-2}$ closed loops. The total number of solution branches $C_{N}=2^{\,N-1}$ partition evenly into $C_{N}/2$ positive and $C_{N}/2$ negative inertial mass branches. The maximum velocity is controlled by the largest bright component particle number.

\section{conclusion and discussion}\label{conclusion}
We report varied branches of nondegenerate vector solitons in multi-component BECs and elucidate their dispersion relations. In the three-component Manakov model, two cubic width constraints generate four admissible branches at a fixed velocity, organizing into two disjoint excitation energy velocity dispersion loops with coexisting positive and negative inertial masses and a common maximum velocity. By increasing the larger bright component particle number, one can decrease both the dispersion relation loops size and the maximal velocity. Both of Bogoliubov-de Gennes analysis and numerical simulations shows that all branches are stable. Under weak linear forces, nondegenerate solitons exhibit dynamics characteristic of positive or negative inertial mass, depending on the branch. Since integrable system forbids conversion between positive and negative mass branches, the nondegenerate soliton fissions into two dark-bright solitons once it is accelerated to its maximal velocity. The highest excitation energy branch fissions into two negative mass DB solitons, whereas the lowest excitation energy branch fissions into two positive mass DB solitons, while the intermediate branches rapidly convert into one positive mass and one negative mass dark-bright soliton. Extending to $N$ components yields $2^{N-1}$ branches in total, comprising $2^{N-2}$ positive mass branches and $2^{N-2}$ negative mass branches. Each a positive mass soliton and a negative mass soliton form a dispersion relation loop, so that the nondegenerate vector soliton's dispersion relation consists of $2^{N-2}$ disjoint loops.

Our results reveal varied branches for nondegenerate solitons comprising one dark component and multiple bright components, originating from the intercomponent coupling. This results suggests that other classes of nondegenerate localized waves may likewise exhibit abundant branches. For example, in multicomponent settings with unequal background densities for the dark components, hence different sound velocity ceilings, distinct maximal velocities are expected, opening the possibility of additional branch splittings. Beyond solitons, our correspondence framework indicates that nondegenerate rogue waves~\cite{nrw1,nrw2,nrw3} and breathers~\cite{ll2016,nkm,nab} should also possess distinct branches. Clarifying the existence conditions, stability properties for nondegenerate localized waves constitutes a natural direction for future work.

Moreover, since our analysis is confined to an integrable system, nondegenerate solitons do not undergo positive negative mass conversion once they reach the maximal velocity, in close analogy with the two-component dark-bright solitons~\cite{gx2025}. Therefore, it would be highly interesting to investigate the dynamics of nondegenerate soliton solutions in nonintegrable models, where such mass conversion may become possible~\cite{zhao2020,Jos2023,mlz2025}.

\section*{Acknowledgments}
L.-C. Zhao was supported by the National Natural Science Foundation of China (Contract No. 12375005, 12235007, 12247103). L. Duan was supported by the National Natural Science Foundation of China (Grant No. 12405003). Y.-H. Qin was supported by the National Natural Science Foundation of China (Grant No. 12405004), the Natural Science Foun dation of Xinjiang Uygur Autonomous Region Project (Grant No. 2024D01C232), the Scientific Research Projects Funded by the Basic Research Business Expenses of Autonomous Region Universities (Grant No. XJEDU2024P011), and the Program of ``Tianchi Talent'' Introduction Plan in Xinjiang Uygur Autonomous Region.


\begin{thebibliography}{99}
\bibitem{soliton4} J. J. Chang, P. Engels, and M. A. Hoefer, Formation of dispersive shock waves by merging and splitting Bose-Einstein condensates,
\newblock\href{https://doi.org/10.1103/PhysRevLett.101.170404}{Phys. Rev. Lett. {\bf101}, 170404 (2008)}.

\bibitem{vs9}
C. Becker, S. Stellmer, P. Soltan-Panahi, S. D\"orscher, M. Baumert, E.-M. Richter, J. Kronj\"ager, K. Bongs, and K. Sengstock,
Oscillations and interactions of dark and dark-bright solitons in Bose-Einstein condensates,
\newblock\href{https://doi.org/10.1038/nphys962}{Nat. Phys. \textbf{4}, 496-501 (2008)}.

\bibitem{vs10}
C. Hamner, J. J. Chang, P. Engels, and M. A. Hoefer,
Generation of dark-bright soliton trains in superfluid-superfluid counterflow,
\newblock\href{https://doi.org/10.1103/PhysRevLett.106.065302}{Phys. Rev. Lett. \textbf{106}, 065302 (2011)}.

\bibitem{ds1}
T. Busch and J. R. Anglin,
Motion of dark solitons in trapped Bose-Einstein condensates,
\newblock\href{https://doi.org/10.1103/PhysRevLett.84.2298}{Phys. Rev. Lett. \textbf{84}, 2298 (2000)}.

\bibitem{ds2}
A. E. Muryshev, G. V. Shlyapnikov, W. Ertmer, K. Sengstock, and M. Lewenstein,
Dynamics of dark solitons in elongated Bose-Einstein condensates,
\newblock\href{https://doi.org/10.1103/PhysRevLett.89.110401}{Phys. Rev. Lett. \textbf{89}, 110401 (2002)}.

\bibitem{bs1} V. E. Zakharov and A. B. Shabat, Exact theory of two-dimensional self-focusing and one-dimensional self-modulation of waves in nonlinear media,
\newblock{Sov. Phys. JETP {\bf34}, 62-69 (1972)}.

\bibitem{bs2} L. Khaykovich, F. Schreck, G. Ferrari, T. Bourdel, J. Cubizolles, L. D. Carr, Y. Castin, and C. Salomon, Formation of a matter-wave bright soliton,
\newblock\href{https://doi.org/10.1126/science.1071021}{Science {\bf296}, 1290-1293 (2002)}.

\bibitem{ds3}
V. A. Brazhnyi, V. V. Konotop, and L. P. Pitaevskii,
Dark solitons as quasiparticles in trapped condensates,
\newblock\href{https://doi.org/10.1103/PhysRevA.73.053601}{Phys. Rev. A \textbf{73}, 053601 (2006)}.

\bibitem{ds4}
L.-Z. Meng, L.-C. Zhao, Th. Busch, and Y. Zhang,
Controlling dark solitons on the healing length scale,
\newblock\href{https://doi.org/10.1088/1361-6455/ad5895}{J. Phys. B: At. Mol. Opt. Phys. \textbf{57}, 145302 (2024)}.

\bibitem{vs5}
H. E. Nistazakis, D. J. Frantzeskakis, P. G. Kevrekidis, B. A. Malomed, and R. Carretero-Gonz{\'a}lez,
Bright-dark soliton complexes in spinor Bose-Einstein condensates,
\newblock\href{https://doi.org/10.1103/PhysRevA.77.033612}{Phys. Rev. A \textbf{77}, 033612 (2008)}.

\bibitem{vs6}
M. Vijayajayanthi, T. Kanna, and M. Lakshmanan,
Bright-dark solitons and their collisions in mixed $N$-coupled nonlinear Schr{\"o}dinger equations,
\newblock\href{https://doi.org/10.1103/PhysRevA.77.013820}{Phys. Rev. A \textbf{77}, 013820 (2008)}.

\bibitem{vs8}
T. Busch and J. R. Anglin,
Dark-bright solitons in inhomogeneous Bose-Einstein condensates,
\newblock\href{https://doi.org/10.1103/PhysRevLett.87.010401}{Phys. Rev. Lett. \textbf{87}, 010401 (2001)}.

\bibitem{vs11}
S. Middelkamp, J. J. Chang, C. Hamner, R. Carretero-Gonz\'alez, P. G. Kevrekidis, V. Achilleos, D. J. Frantzeskakis, P. Schmelcher, and P. Engels,
Dynamics of dark-bright solitons in cigar-shaped Bose-Einstein condensates,
\newblock\href{https://doi.org/10.1016/j.physleta.2010.11.025}{Phys. Lett. A \textbf{375}, 642-646 (2011)}.

\bibitem{vs12}
D. Yan, J. J. Chang, C. Hamner, P. G. Kevrekidis, P. Engels, V. Achilleos, D. J. Frantzeskakis, R. Carretero-Gonz\'alez, and P. Schmelcher,
Multiple dark-bright solitons in atomic Bose-Einstein condensates,
\newblock\href{https://doi.org/10.1103/PhysRevA.84.053630}{Phys. Rev. A \textbf{84}, 053630 (2011)}.

\bibitem{bd} Y.-H. Qin, Y. Wu, L.-C. Zhao, and Z.-Y. Yang, Interference properties of two-component matter wave solitons,
\newblock\href{https://doi.org/10.1088/1674-1056/ab65b7}{Chin. Phys. B {\bf29}, 020303 (2020)}.

\bibitem{zhao2020}
L.-C. Zhao, W. Wang, Q. Tang, Z.-Y. Yang, W.-L. Yang, and J. Liu,
Spin soliton with a negative-positive mass transition,
\newblock\href{https://doi.org/10.1103/PhysRevA.101.043621}{Phys. Rev. A \textbf{101}, 043621 (2020)}.

\bibitem{Jos2023}
S. Bresolin, A. Roy, G. Ferrari, A. Recati, and N. Pavloff,
Oscillating solitons and ac Josephson effect in ferromagnetic Bose-Bose mixtures,
\newblock\href{https://doi.org/10.1103/PhysRevLett.130.220403}{Phys. Rev. Lett. \textbf{130}, 220403 (2023)}.

\bibitem{mlz2025}
L.-Z. Meng, X.-W. Luo, and L.-C. Zhao,
Self-adapted Josephson oscillation of dark-bright solitons under constant forces,
\newblock\href{https://doi.org/10.1103/8l2y-d7cd}{Phys. Rev. A \textbf{112}, 033306 (2025)}.

\bibitem{manakov}
S. V. Manakov,
On the theory of two-dimensional stationary self-focusing of electromagnetic waves,
Sov. Phys. JETP \textbf{38}, 248 (1974).

\bibitem{gx2025}
X. Gao, L.-Z. Meng, and L.-C. Zhao,
Dark-bright solitons with positive mass in Manakov cases with repulsive interactions,
\newblock\href{https://doi.org/10.1103/PhysRevE.111.054209}{Phys. Rev. E \textbf{111}, 054209 (2025)}.

\bibitem{dbb1} T. M. Bersano, V. Gokhroo, M. A. Khamehchi, J. D\'Ambroise, D. J. Frantzeskakis, P. Engels, and P. G. Kevrekidis, Three-component soliton states in spinor $F=1$ Bose-Einstein condensates,
\newblock\href{https://doi.org/10.1103/PhysRevLett.120.063202}{Phys. Rev. Lett. {\bf120}, 063202 (2018)}.

\bibitem{vs16}
B.-F. Feng,
General $N$-soliton solution to a vector nonlinear Schr\"odinger equation,
\newblock\href{https://doi.org/10.1088/1751-8113/47/35/355203}{J. Phys. A: Math. Theor. \textbf{47}, 355203 (2014)}.

\bibitem{vs3}
S. Stalin, R. Ramakrishnan, M. Senthilvelan, and M. Lakshmanan,
Nondegenerate solitons in Manakov system,
\newblock\href{https://doi.org/10.1103/PhysRevLett.122.043901}{Phys. Rev. Lett. \textbf{122}, 043901 (2019)}.

\bibitem{vs31} R. Ramakrishnan, S. Stalin, and M. Lakshmanan, Nondegenerate solitons and their collisions in Manakov systems,
\newblock\href{https://doi.org/10.1103/PhysRevE.102.042212}{Phys. Rev. E {\bf102}, 042212 (2020)}.

\bibitem{vs4}
Y.-H. Qin, L.-C. Zhao, and L. Ling,
Nondegenerate bound-state solitons in multicomponent Bose-Einstein condensates,
\newblock\href{https://doi.org/10.1103/PhysRevE.100.022212}{Phys. Rev. E \textbf{100}, 022212 (2019)}.

\bibitem{ndbb}
Y.-H. Qin, L.-C. Zhao, Z.-Q. Yang, and L. Ling,
Multivalley dark solitons in multicomponent Bose-Einstein condensates with repulsive interactions,
\newblock\href{https://doi.org/10.1103/PhysRevE.104.014201}{Phys. Rev. E \textbf{104}, 014201 (2021)}.

\bibitem{ns1}
S. Stalin, R. Ramakrishnan, and M. Lakshmanan,
Nondegenerate bright solitons in coupled nonlinear Schr\"odinger systems: Recent developments on optical vector solitons,
\newblock\href{https://doi.org/10.3390/photonics8070258}{Photonics \textbf{8}, 258 (2021)}.

\bibitem{ns2}
J. Rao, D. Mihalache, J. He, and F. Zhou,
Degenerate and non-degenerate vector solitons and their interactions in the two-component long-wave-short-wave model of Newell type,
\newblock\href{https://doi.org/10.1016/j.chaos.2022.112963}{Chaos Solitons Fractals \textbf{166}, 112963 (2023)}.

\bibitem{ns21} K.-L. Geng, D.-S. Mou, and C.-Q. Dai, Nondegenerate solitons of 2-coupled mixed derivative nonlinear Schr\"odinger equations,
\newblock\href{https://doi.org/10.1007/s11071-022-07833-5}{Nonlinear Dyn. {\bf111}, 603-617 (2023)}.

\bibitem{ns22} Y.-J. Cai, J.-W. Wu, and J. Lin, Nondegenerate N-soliton solutions for Manakov system,
\newblock\href{https://doi.org/10.1016/j.chaos.2022.112657}{Chaos, Solitons Fractals {\bf164}, 112657 (2022)}.

\bibitem{ns23} W.-X. Qiu, Z.-Z. Si, D.-S. Mou, C.-Q. Dai, J.-T. Li, and W. Liu, Data-driven vector degenerate and nondegenerate solitons of coupled nonlocal nonlinear Schr\"odinger equation via improved PINN algorithm,
\newblock\href{https://doi.org/10.1007/s11071-024-09648-y}{Nonlinear Dyn. {\bf113}, 4063-4076 (2025)}.

\bibitem{vs7}
L.-C. Zhao, Z.-Y. Yang, and W.-L. Yang,
Solitons in nonlinear systems and eigen-states in quantum wells,
\newblock\href{https://doi.org/10.1088/1674-1056/28/1/010501}{Chin. Phys. B \textbf{28}, 010501 (2019)}.

\bibitem{eff1}
L. D. Landau and E. M. Lifshitz,
\textit{Quantum Mechanics: Non-Relativistic Theory},
\newblock{Pergamon, Oxford (1977)}.

\bibitem{gp1}
B. A. Malomed,
\textit{Emergent Nonlinear Phenomena in Bose-Einstein Condensates: Theory and Experiment},
\newblock{Springer, Berlin (2008)}.

\bibitem{gp2}
P. G. Kevrekidis and D. J. Frantzeskakis,
Solitons in coupled nonlinear Schr\"odinger models: A survey of recent developments,
\newblock\href{https://doi.org/10.1016/j.revip.2016.07.002}{Rev. Phys. \textbf{1}, 140-153 (2016)}.

\bibitem{gp3}
D.-S. Wang, W. Han, Y.-R. Shi, Z.-D. Li, and W.-M. Liu,
Dynamics and stability of stationary states for the spin-1 Bose-Einstein condensates in a standing light wave,
\newblock\href{https://doi.org/10.1016/j.cnsns.2015.11.018}{Commun. Nonlinear Sci. Numer. Simul. \textbf{36}, 45-57 (2016)}.

\bibitem{darb1}
C. Rogers and W. K. Schief,
\textit{B\"acklund and Darboux Transformations: Geometry and Modern Applications in Soliton Theory},
\newblock{Cambridge University Press, Cambridge (2002)}.

\bibitem{darb2}
E. V. Doktorov and S. B. Leble,
\textit{A Dressing Method in Mathematical Physics},
\newblock{Springer, Dordrecht (2007)}.

\bibitem{hb}
R. Hirota,
\textit{The Direct Method in Soliton Theory},
\newblock{Cambridge University Press, Cambridge (2004)}.

\bibitem{invs}
S. P. Novikov, S. V. Manakov, L. P. Pitaevskii, and V. E. Zakharov,
\textit{Theory of Solitons: The Inverse Scattering Method},
\newblock{Consultants Bureau, New York (1984)}.

\bibitem{vs1}
X.-F. Zhang, X.-H. Hu, X.-X. Liu, and W. M. Liu,
Vector solitons in two-component Bose-Einstein condensates with tunable interactions and harmonic potential,
\newblock\href{https://doi.org/10.1103/PhysRevA.79.033630}{Phys. Rev. A \textbf{79}, 033630 (2009)}.

\bibitem{vs2}
T. Kanna and M. Lakshmanan,
Exact soliton solutions, shape changing collisions, and partially coherent solitons in coupled nonlinear Schr{\"o}dinger equations,
\newblock\href{https://doi.org/10.1103/PhysRevLett.86.5043}{Phys. Rev. Lett. \textbf{86}, 5043 (2001)}.

\bibitem{vs14}
L. Ling, L.-C. Zhao, and B. Guo,
Darboux transformation and multi-dark soliton for $N$-component nonlinear Schr\"odinger equations,
\newblock\href{https://doi.org/10.1088/0951-7715/28/9/3243}{Nonlinearity \textbf{28}, 3243-3261 (2015)}.

\bibitem{vs15}
Y. Ohta, D.-S. Wang, and J. Yang,
General $N$-dark-dark solitons in the coupled nonlinear Schr\"odinger equations,
\newblock\href{https://doi.org/10.1111/j.1467-9590.2011.00525.x}{Stud. Appl. Math. \textbf{127}, 345-371 (2011)}.

\bibitem{mi} K. Tai, A. Hasegawa, and A. Tomita, Observation of modulational instability in optical fibers,
\newblock\href{https://doi.org/10.1103/PhysRevLett.56.135}{Phys. Rev. Lett. {\bf56}, 135-138 (1986)}.

\bibitem{es1}
V. A. Brazhnyi and V. M. P\'erez-Garc\'ia,
Stable multidimensional soliton stripes in two-component Bose-Einstein condensates,
\newblock\href{https://doi.org/10.1016/j.chaos.2010.12.012}{Chaos Solitons Fractals \textbf{44}, 381-389 (2011)}.

\bibitem{es2}
R. G. Scott, F. Dalfovo, L. P. Pitaevskii, and S. Stringari,
Dynamics of dark solitons in a trapped superfluid Fermi gas,
\newblock\href{https://doi.org/10.1103/PhysRevLett.106.185301}{Phys. Rev. Lett. \textbf{106}, 185301 (2011)}.

\bibitem{bdtrap2}
E. T. Karamatskos, J. Stockhofe, P. G. Kevrekidis and P. Schmelcher. Stability and tunneling dynamics of a dark-bright soliton pair in a harmonic trap,
\newblock\href{https://journals.aps.org/pra/abstract/10.1103/PhysRevA.91.043637}
{Phys. Rev. A \textbf{91}, 043637(2015)}.

\bibitem{bdtrap3}
Majed O. D. Alotaibi and Lincoln D. Carr. Internal oscillations of a dark-bright soliton in a harmonic potential,
\newblock\href{https://iopscience.iop.org/article/10.1088/1361-6455/aadfb2}
{J. Phys. B: At. Mol. Opt. Phys. \textbf{51} 205004(2018)}.

\bibitem{wbh2024}
B.-H Wang, N Mao, L.-C. Zhao,
The complete set of eigenstates in one type of N-multiple quantum wells,
\newblock\href{https://doi.org/10.1088/1402-4896/ad21cb}{Phys. Scr.\textbf{99}, 035108 (2024).}

\bibitem{nrw1} L. Ling, B. Guo, and L.-C. Zhao, High-order rogue waves in vector nonlinear Schr\"odinger equations,
\newblock\href{https://doi.org/10.1103/PhysRevE.89.041201}{Phys. Rev. E {\bf89}, 041201 (2014)}.

\bibitem{nrw2} L.-C. Zhao, B. Guo, and L. Ling, High-order rogue wave solutions for the coupled nonlinear Schr\"odinger equations-II,
\newblock\href{https://doi.org/10.1063/1.4947113}{J. Math. Phys. {\bf57}, 043508 (2016)}.

\bibitem{nrw3}
C. Liu, S.-C. Chen, X. Yao, and N. Akhmediev,
Non-degenerate multi-rogue waves and easy ways of their excitation,
\newblock\href{https://doi.org/10.1016/j.physd.2022.133192}{Physica D \textbf{433}, 133192 (2022)}.

\bibitem{ll2016} L. Ling and L.-C. Zhao, Modulational instability and homoclinic orbit solutions in vector nonlinear Schr\"odinger equation,
\newblock\href{https://doi.org/10.1016/j.cnsns.2019.01.008}{Commun. Nonlinear Sci. Numer. Simul. {\bf72}, 449-471 (2019)}.

\bibitem{nkm} W.-J. Che, S.-C. Chen, C. Liu, L.-C. Zhao, and N. Akhmediev, Nondegenerate Kuznetsov-Ma solitons of Manakov equations and their physical spectra,
\newblock\href{https://doi.org/10.1103/PhysRevA.105.043526}{Phys. Rev. A {\bf105}, 043526 (2022)}.

\bibitem{nab} C. Liu, S.-C. Chen, X. Yao, and N. Akhmediev, Modulation instability and non-degenerate Akhmediev breathers of Manakov equations,
\newblock\href{https://doi.org/10.1088/0256-307X/39/9/094201}{Chin. Phys. Lett. {\bf39}, 094201 (2022)}.

\end{thebibliography}
\end{document}